\documentclass[11pt,preprint]{aastex}

\usepackage{epsfig}
\usepackage{graphicx}
\usepackage{url}
\usepackage{natbib}
\usepackage{float}

\shorttitle{Kinematic substructure in the disk}
\shortauthors{Carlin, DeLaunay, et al.}

\begin{document}

\title{Substructure in bulk velocities of Milky Way disk stars}

\author{
Jeffrey L. Carlin\altaffilmark{1,2}, 
James DeLaunay\altaffilmark{1,2}, 
Heidi Jo Newberg\altaffilmark{2}, 
Licai Deng\altaffilmark{3},  
Daniel Gole\altaffilmark{2,4}, 
Kathleen Grabowski\altaffilmark{2}, 
Ge Jin\altaffilmark{5},
Chao Liu\altaffilmark{3},
Xiaowei Liu\altaffilmark{6},
A-Li Luo\altaffilmark{3},
Haibo Yuan\altaffilmark{6},
Haotong Zhang\altaffilmark{3},
Gang Zhao\altaffilmark{3},
Yongheng Zhao\altaffilmark{3}
}

\altaffiltext{1}{Equal first authors.}
\altaffiltext{2}{Department of Physics, Applied Physics and Astronomy, Rensselaer Polytechnic Institute, Troy, NY 12180, USA, carlij@rpi.edu}
\altaffiltext{3}{Key Lab for Optical Astronomy, National Astronomical Observatories, Chinese Academy of Sciences, Beijing 100012, China}
\altaffiltext{4}{Department of Physics, SUNY-Geneseo}
\altaffiltext{5}{University of Science and Technology of China, Hefei 230026, China}
\altaffiltext{6}{Kavli Institute for Astronomy and Astrophysics, Peking University, Beijing 100871, China; Department of Astronomy, Peking University, Beijing 100871, China}

\begin{abstract}

We find that Galactic disk stars near the anticenter exhibit velocity asymmetries in both the Galactocentric radial and vertical components across the mid-plane as well as azimuthally. These findings are based on LAMOST spectroscopic velocities for a sample of $\sim400,000$~F-type stars, combined with proper 
motions from the PPMXL catalog for which we have derived corrections to the zero points based in part on spectroscopically discovered galaxies and 
QSOs from LAMOST. In the region within 2 kpc outside the Sun's radius and $\pm2$~kpc from the Galactic midplane, we show that stars above the plane 
exhibit net outward radial motions with downward vertical velocities, while stars below the plane have roughly the opposite behavior. We discuss this in 
the context of other recent findings, and conclude that we are likely seeing the signature of vertical disturbances to the disk due to an external 
perturbation.

\end{abstract}


\section{Introduction}

A variety of instabilities can produce non-axisymmetric features near the midplane of the Galaxy, including 
resonances due to the bar and/or spiral arms (e.g., \citealt{f01,qm05,avp+09,mqw+09,mbs+10,qdb+11}), and (possibly associated) processes such as radial migration (e.g., \citealt{sb02, h08a, mf10}). In contrast, any {\it vertical} structures that are seen must have been excited by an external means such as a passing satellite galaxy or some other halo 
substructure. Vertical density and velocity structures have been shown to arise in model Milky Way disks due to perturbation by a Sagittarius (Sgr) sized 
dwarf galaxy \citep[see also \citealt{pbt+11}]{gmo+13}. \citet{wgy+12} showed the existence of vertical, wavelike 
structures in stellar density from Sloan Digital Sky Survey (SDSS) data. Stars in this structure apparently also exhibit vertical motions (perpendicular to 
the disk), suggesting that the structures are coherent perturbations in the disk. \citet{wgy+12} modeled the effect of a massive perturber on the vertical 
structure of the Galactic disk, and showed that fairly short-lived vertical waves are produced in such encounters. Analytically, these authors showed that 
vertical perturbations should have wavelengths slightly less than 2 kpc. \citet{gmo+12} showed that peaks in the energies of SDSS-selected thick disk 
stars are consistent with their predictions of merger-induced waves in the disk, which can be long-lived if the perturber is sufficiently massive.

Velocity structure in both the vertical and radial directions was also seen in a sample of RAVE (RAdial Velocity Experiment; \citealt{szs+06}) 
stars within 2~kpc of the Sun by \citet{wsb+13}. This study probed mostly the fourth Galactic quadrant, and thus contains the majority of its data inside the 
solar radius. Nonetheless, apparently similar wavelike structures are seen in the different volumes probed by the RAVE \citep{wsb+13} and SDSS 
\citep{wgy+12} studies.

In this Letter, we extend the work of these previous studies using a much larger, independent data set. We use spectra from the LAMOST survey 
combined with corrected proper motions from the PPMXL catalog to derive three-dimensional space velocities for a large sample of stars with known 
spectral types. From our sample of $\sim400,000$ F-type stars near the Galactic anticenter, we demonstrate the existence of a kinematic asymmetry 
above and below the Galactic plane in the 2~kpc cylindrical shell just outside of the Sun's radius. The asymmetry is present in both the radial and vertical 
components of the Galactocentric velocities.

\section{Data}
\subsection{LAMOST spectra}\label{sec:lamost_spectra}

As of June 2013, the LAMOST (Large Sky Area Multi-Object Fiber Spectroscopic Telescope: \citealt{czc+12, zzc+12}) survey has completed its first 
season of official operations, in addition to a year of pilot survey data collection. In these first two observing seasons, $\sim1.7$~million spectra have 
been obtained that have sufficient signal to noise ($S/N$) for velocity/redshift measurements. The majority of these stars were observed 
as part of the Milky Way structure portion of the survey, known as LEGUE (LAMOST Experiment for Galactic Understanding and Exploration; 
\citealt{dnl+12}). LAMOST observes at a resolution of $R \equiv \frac{\Delta \lambda}{\lambda} \sim 1800$ over a wavelength range of $3800 \lesssim 
\lambda \lesssim 9100$~\AA, which produces spectra similar to those from SDSS \citep{yrn+09}.

From the $\sim1.7$~million spectra available to the collaboration, $\sim1.3$~million are stellar spectra with $S/N > 5$ in the 
SDSS $g, r$, and $i$ bands. Although the process for determination of stellar parameters (i.e., $\log g$, $T_{\rm eff}$, and [Fe/H]) from LAMOST spectra is still being 
refined, the radial velocities (RVs) and spectral types are well measured (see, e.g., \citealt{lzz+12}). The spectra are mostly of bright ($r \lesssim 17$) stars 
spanning a large range of photometric colors (for more information about LEGUE pilot survey target selection, see \citealt{cln+12, chy+12, ycl+12, zcy+12}, and for a description of the LAMOST Galactic anticenter survey and the Xuyi photometric survey from which its targets were selected, see \citealt{lyh+13}). Such a sample will inevitably contain a majority of nearby main sequence stars, since these dominate the number density along any line of sight. 

\begin{table}[!t]
\caption{Selection criteria for the stellar sample. Ranges in $V_R$, $V_{\theta}$, and $V_Z$ were chosen to exclude high-velocity, nearby halo stars.}
\begin{center}
\begin{tabular}{c}

\tableline 

classified as F-type star\\
$S/N > 5$ in $g, r$, and $i$\\
error in $V_{\rm helio} <$ 10 km s$^{-1}$\\
error in $\mu_{\alpha} \cos \delta$, $\mu_{\delta} < 10$ mas yr$^{-1}$\\
$-150 < V_R < 150$ km s$^{-1}$\\
$-400 < V_{\theta} < -100$ km s$^{-1}$\\
$-150 < V_Z < 150$ km s$^{-1}$\\
$7.8 < R_{\rm GC} < 9.8$ kpc\\
$-2.0 < Z_{\rm GC} < 2.0$ kpc\\

\tableline


\end{tabular}
\end{center}
\label{tab:selection}
\end{table}

For this study, we selected stars classified as F-type by the LAMOST pipeline 
that have 
$J, H,$ and $K_S$ magnitudes from the Two Micron All Sky Survey (2MASS; \citealt{scs+06}). 
Magnitudes were corrected for extinction using E(B-V) derived 
from the maps of \citet{sfd98}, with coefficients for 2MASS bands from \citet{ccm89}. 
Due to site conditions and the location of the Guoshoujing 
telescope, most of the LEGUE data are concentrated near the Galactic anticenter. 
The $S/N$, velocity error, spatial, and velocity criteria used to select the sample are given in Table~\ref{tab:selection}. These criteria yield $380,159$ LAMOST F-stars, and an additional $\sim10,737$ RAVE stars (see Section~\ref{sec:rave}).

We derive distances based on calibration to Hipparcos \citep{plk+97} data with spectroscopic information from \citet{g06c}. We select F-type stars from 
the Gontcharov catalog and remove giants and subdwarfs. 
For stars with reliable parallaxes, we derive 
average absolute magnitudes of $M_{K_S} = 1.57$, 1.83, and 2.18 for F2, F5, and F9 stars, respectively (using $\pm1$ sub-class; e.g., ``F2'' includes F1-F3 stars). Distances to LAMOST F-stars were derived using the absolute magnitude of the 
type nearest the observed LAMOST spectral sub-type. Note that because we used $K_S$-band magnitudes, the effects on our derived distances of overestimating extinction to nearby, low-latitude stars are minimal.

\subsection{Proper motion corrections}\label{sec:pmcorr}

We match the LAMOST data to PPMXL \citep{rds10}. PPMXL proper motions of QSOs are known to contain systematic shifts that vary with 
position on the sky (see, e.g., \citealt{wmz11}). We derive new corrections to the proper motion reference frame using 100,919 QSOs from the catalog of 
\citet{vv10} plus 12,619 QSOs and galaxies spectroscopically identified by LAMOST (and also in PPMXL). After removing objects with large proper motion errors, we fit polynomials in two dimensions to the mean proper 
motions of QSOs (which should be zero) as a function of position on the sky. Coefficients for these fits are given in Table~\ref{tab:pmfits}. 
Our fits show similar behavior to that seen in \citet{wmz11} -- the $\mu_{\alpha} \cos \delta$ residuals are low (i.e., negative) in the north Galactic hemisphere, and near zero elsewhere, and there are systematic shifts of $\sim 2$ mas~yr$^{-1}$ in both proper motion dimensions. 
The $\mu_{\delta}$ fit increases mostly as a function of declination.
In all subsequent discussion, the proper motions were pre-corrected with these polynomial fits before using the data for analysis.

\begin{figure}[!t]
\includegraphics[width=0.49\textwidth, trim=1.0cm 1.0cm 1.0cm 1.0cm, clip]{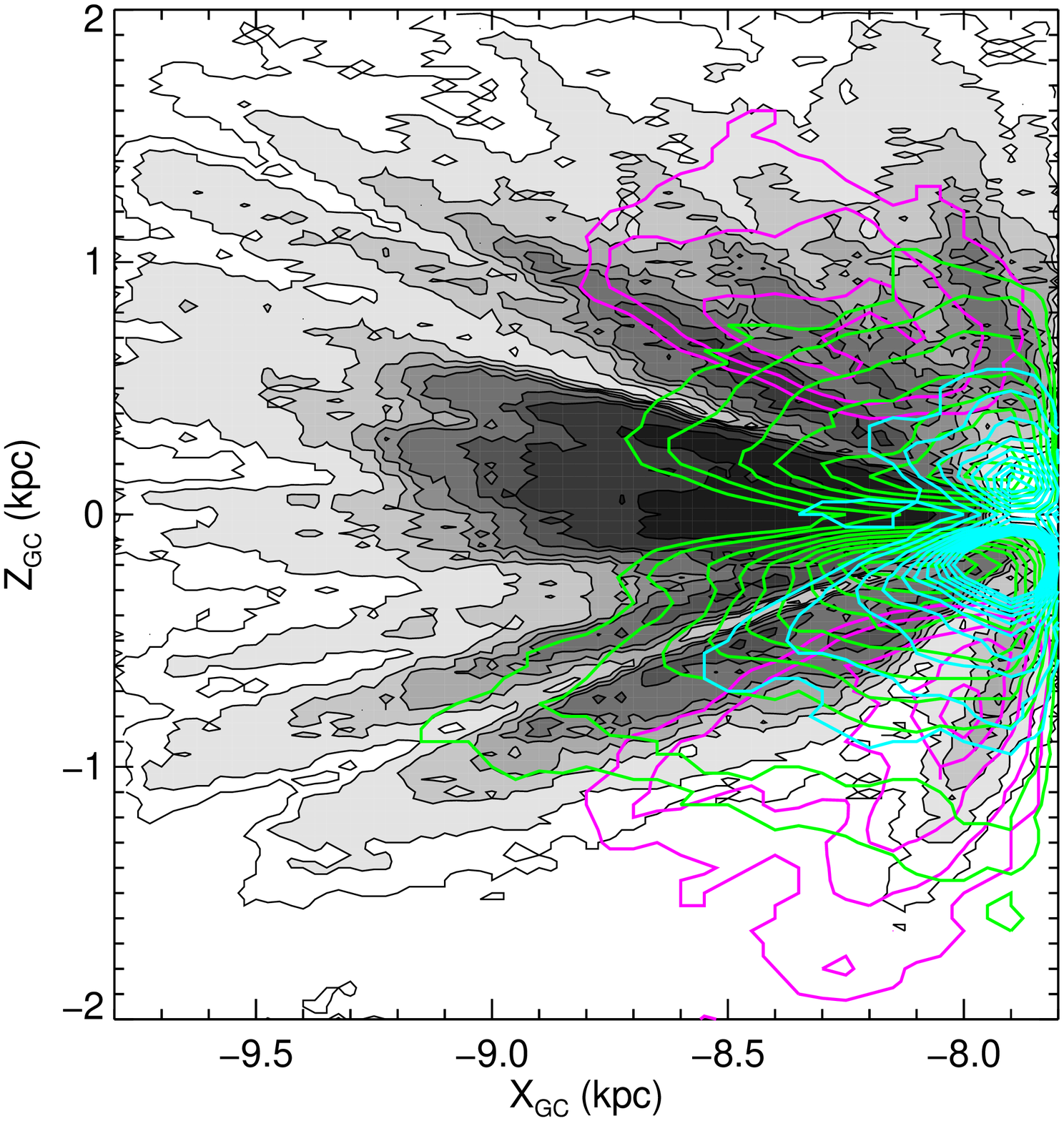}
\includegraphics[width=0.49\textwidth, trim=1.0cm 1.0cm 1.0cm 1.0cm, clip]{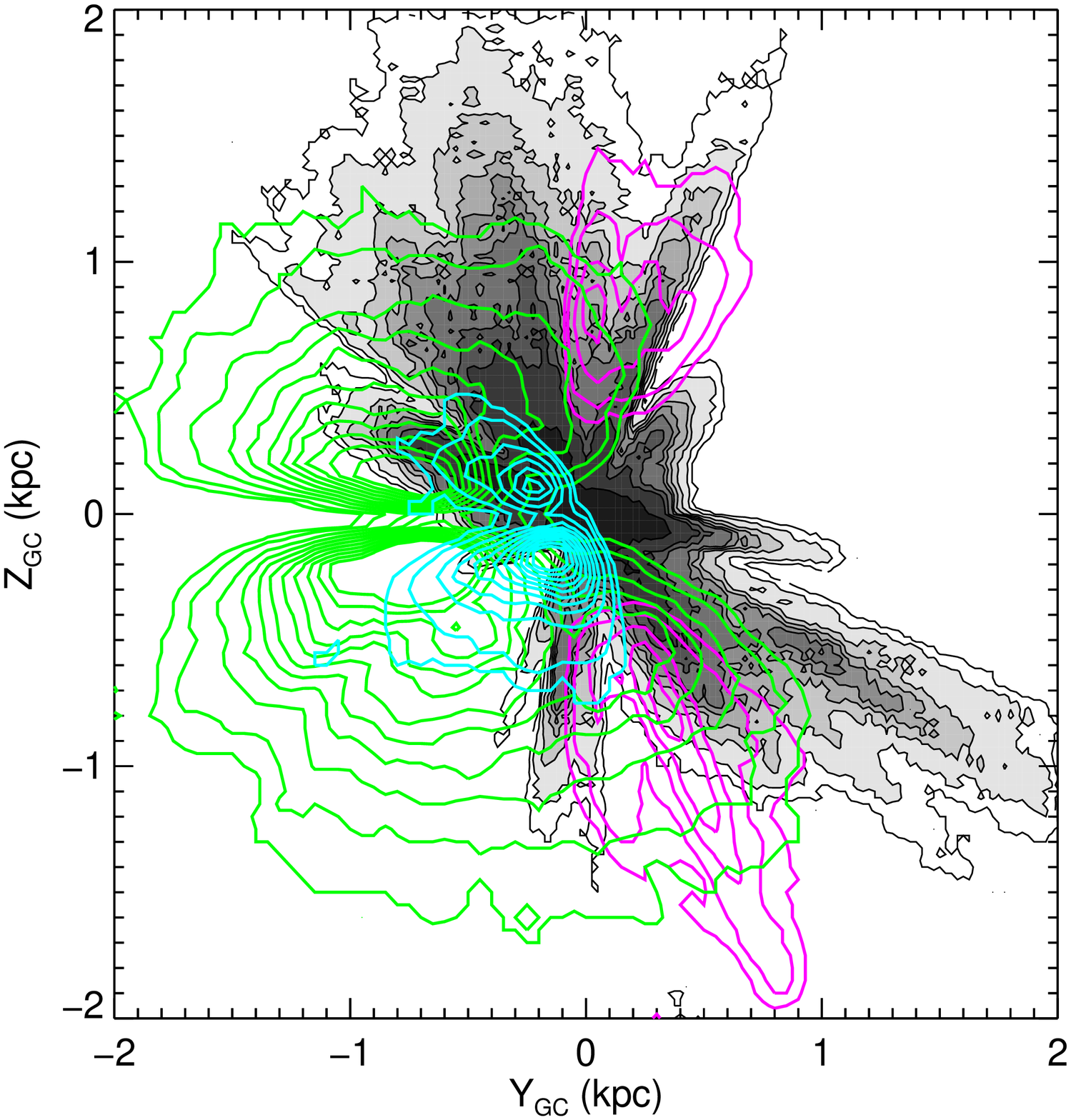}
\includegraphics[width=0.49\textwidth, trim=1.0cm 1.0cm 1.0cm 1.0cm, clip]{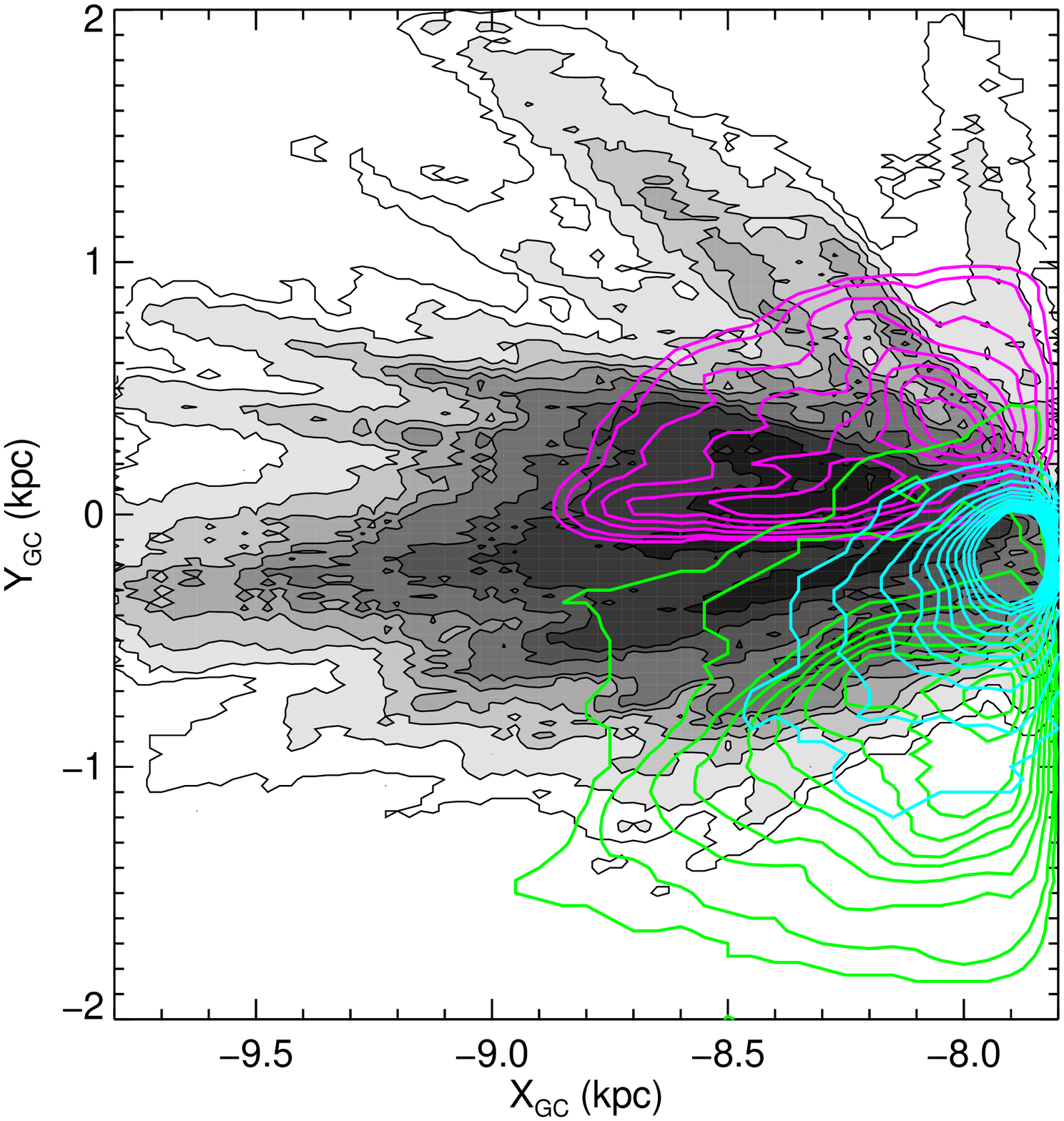}
\includegraphics[width=0.5\textwidth, trim=1.0cm 1.0cm 1.0cm 1.0cm, clip]{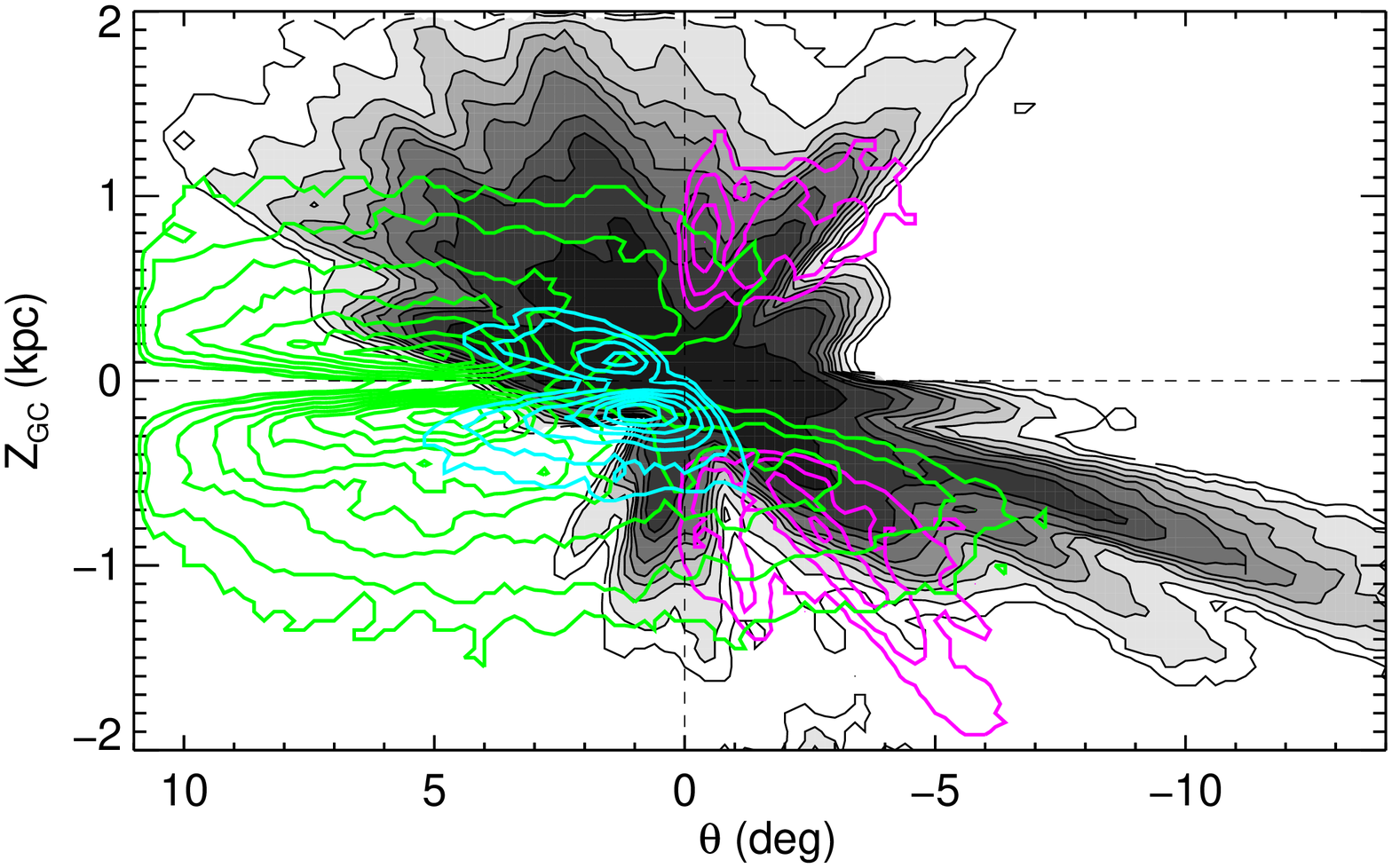}

\caption{Spatial distribution, in Galactic coordinates, of the stars with LAMOST spectra used in this study. The Sun is at $(X,Y,Z)_{\rm GC} = (-7.8, 0, 0)$~kpc. Smoothed, grayscale contours represent the number of stars in 250$\times$250~pc bins, with levels of 5, 10, 20, 30, 40, 50, 75, 100, and 200 stars per bin. Magenta contours show the SDSS data from \citet{wgy+12}, green contours the RAVE red clump sample from \citet{wsb+13}, and cyan contours denote the RAVE sample used in our study.
}
   \label{fig:xyz}
\end{figure}

\begin{table}[!t]
\caption{Proper motion fits as a function of RA, Dec.}
\begin{center}
\begin{tabular}{crrrrrr}

\tableline 
\multicolumn{1}{c}{fit} & \multicolumn{1}{c}{a} & \multicolumn{1}{c}{b} & \multicolumn{1}{c}{c} & \multicolumn{1}{c}{d} & \multicolumn{1}{c}{e} & \multicolumn{1}{c}{f} \\

\tableline
$\mu_{\alpha} \cos \delta$ & 1.8986 &  -4.1637E-02 & 1.0376E-04 &  1.2363E-04 & -5.1269E-02 &  4.0594E-04\\
$\mu_{\delta}$ & -3.2247 &  1.0490E-02 & -2.9027E-05 & 2.1413E-04 & -6.3930E-02 & 1.3776E-03

\end{tabular}
\end{center}
\tablenotetext{}{$\mu_{\rm fit} = a + (b \alpha) + (c \alpha^2) + (d \alpha \delta) + (e \delta) + (f \delta^2)$, where $\mu_{\rm fit}$ is the proper motion being fit, in mas yr$^{-1}$, and $(\alpha, \delta)$ are right ascension and declination, both in degrees.}
\label{tab:pmfits}
\end{table}

\subsection{Coordinate system and velocity calculations}\label{sec:coords}

We derive positions and velocities in Galactic Cartesian ($XYZ_{\rm GC}$) and cylindrical ($R_{\rm GC}, \theta, Z_{\rm GC}$) coordinate systems with origins at the Galactic center. The Cartesian $X_{\rm GC}$-axis is positive toward the Galactic center, with the Sun at $X_{\rm GC} = -7.8$~kpc \citep{mb10}. $Y_{\rm GC}$ is in the direction of Galactic rotation, and $Z_{\rm GC}$ is positive toward the north Galactic pole. The distribution of the LAMOST sample in $XYZ_{\rm GC}$ is shown in Figure~\ref{fig:xyz}. Galactocentric cylindrical coordinates place the Sun at $(R, \theta, Z)_{\rm GC} = (7.8~{\rm kpc}, 0^\circ, 0~{\rm kpc})$, with $\theta$ increasing in the same direction as Galactic longitude.
Three-dimensional Cartesian ($U, V, W$) velocities are corrected to a Galactocentric frame by removing the \citet{sbd10} values for the solar peculiar motion, $(U, V, W)_{\odot} = 
(11.1, 12.24, 7.25)$ km~s$^{-1}$, and a local standard of rest velocity of 247 km~s$^{-1}$ \citep{mb10}. 
Because $V_{\theta}$ 
increases in the direction of Galactic longitude (opposite the direction of rotation), the cylindrical rotation velocity in the solar neighborhood is -247~km~s$^{-1}$.

\subsection{RAVE spectra}\label{sec:rave}

To more completely sample the sky in the volume of the Galaxy we are studying, we added data from the southern-hemisphere RAVE survey Data 
Release 3 \citep{sws+11} to our sample. Using distances given by the RAVE catalog (from \citealt{zmb+10}), we select stars from the same volume of the 
Milky Way as the LAMOST sample (see Table~\ref{tab:selection} for selection criteria). This adds 10,737 stars, most of which are in a small volume of the third 
Galactic quadrant just below the midplane. The addition of the publicly available RAVE data acts as confirmation of the LAMOST results in the third quadrant.

\section{Velocity asymmetries among nearby disk stars}

\begin{figure}[!t]
\begin{center}
 \includegraphics[width=0.49\textwidth, trim=0.5cm 0.0cm 0.5cm 0.5cm, clip]{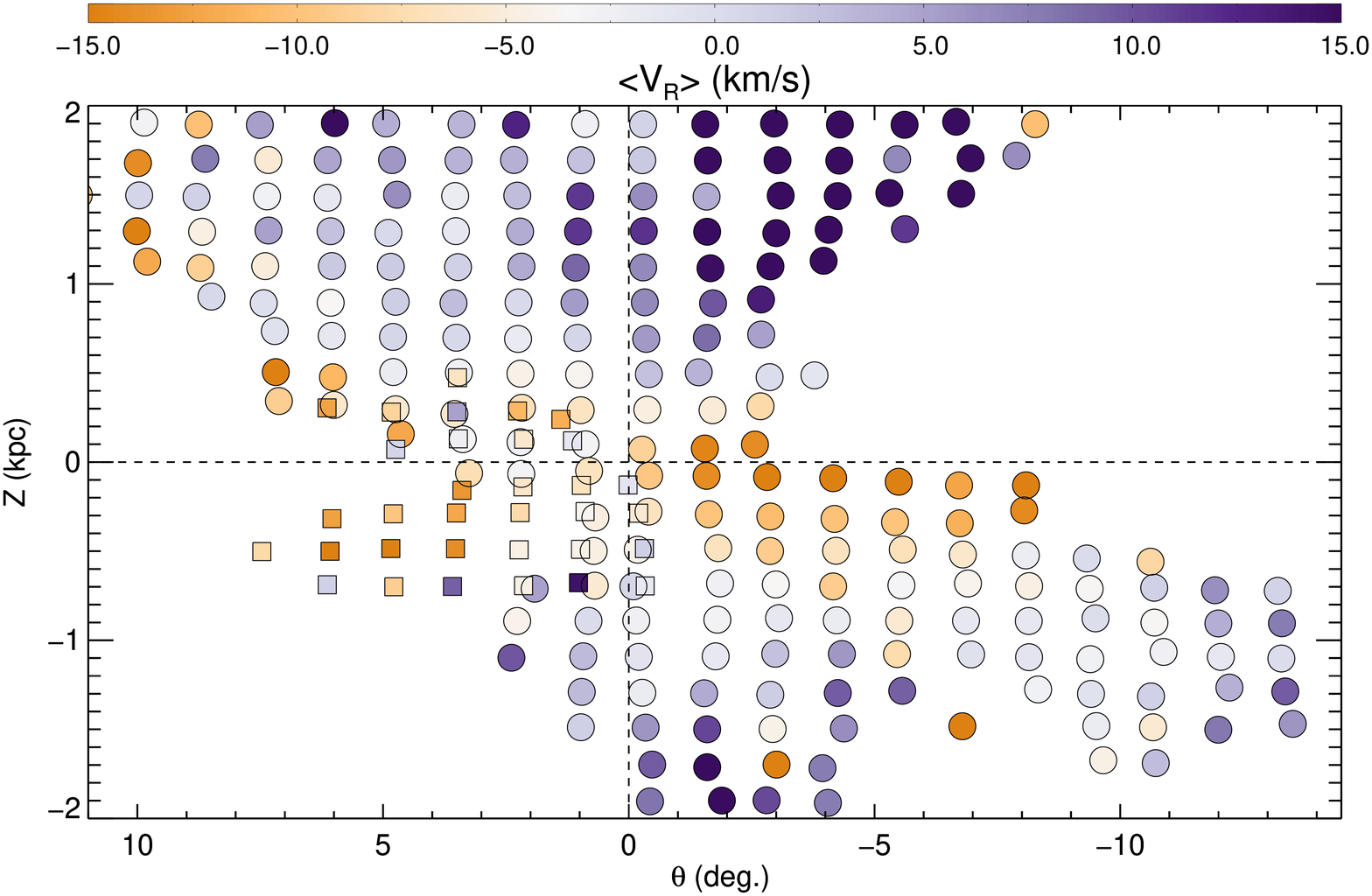}
 \includegraphics[width=0.49\textwidth, trim=0.5cm 0.0cm 0.5cm 0.5cm, clip]{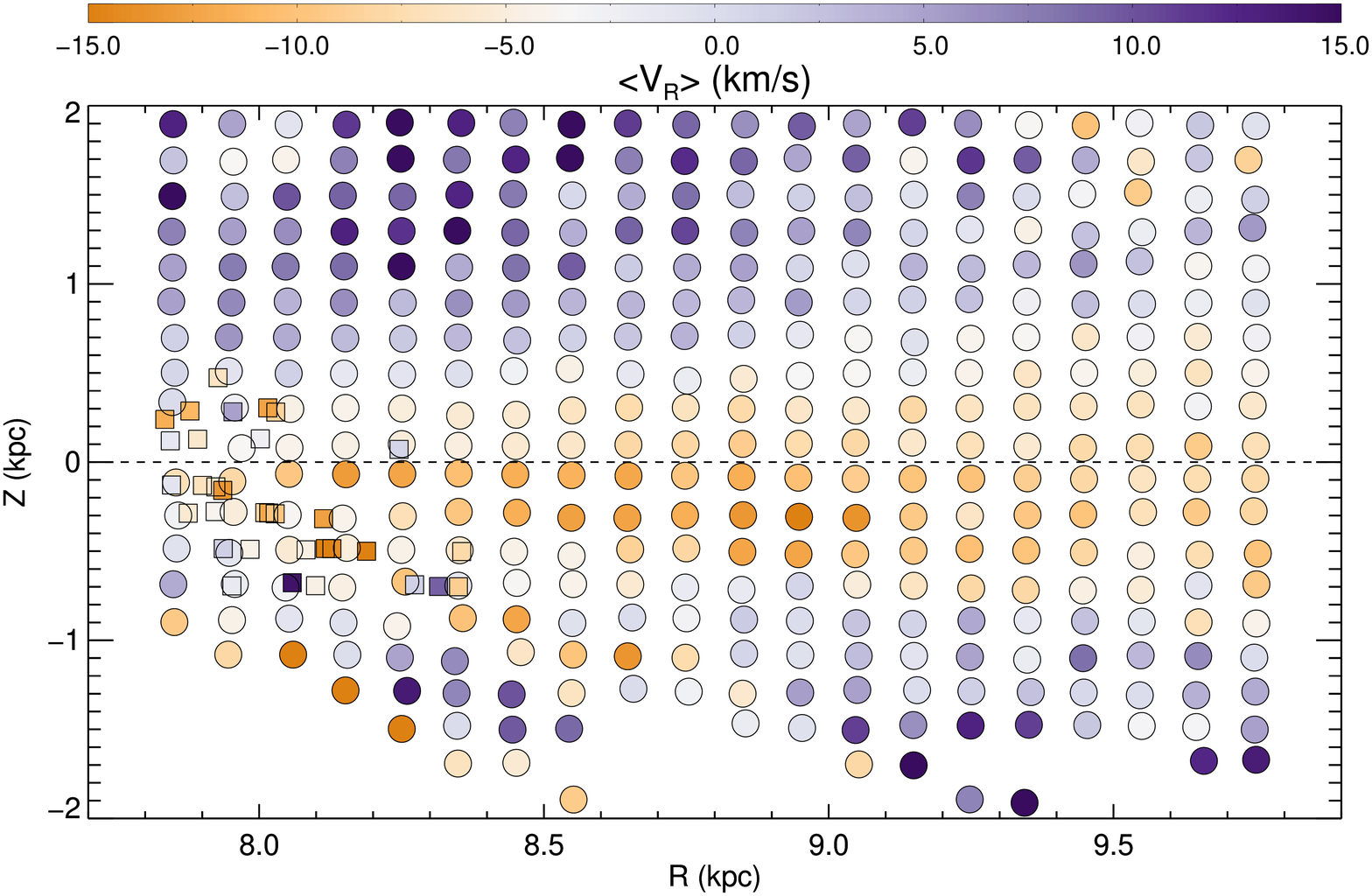}
 \includegraphics[width=0.49\textwidth, trim=0.5cm 0.0cm 0.5cm 0.5cm, clip]{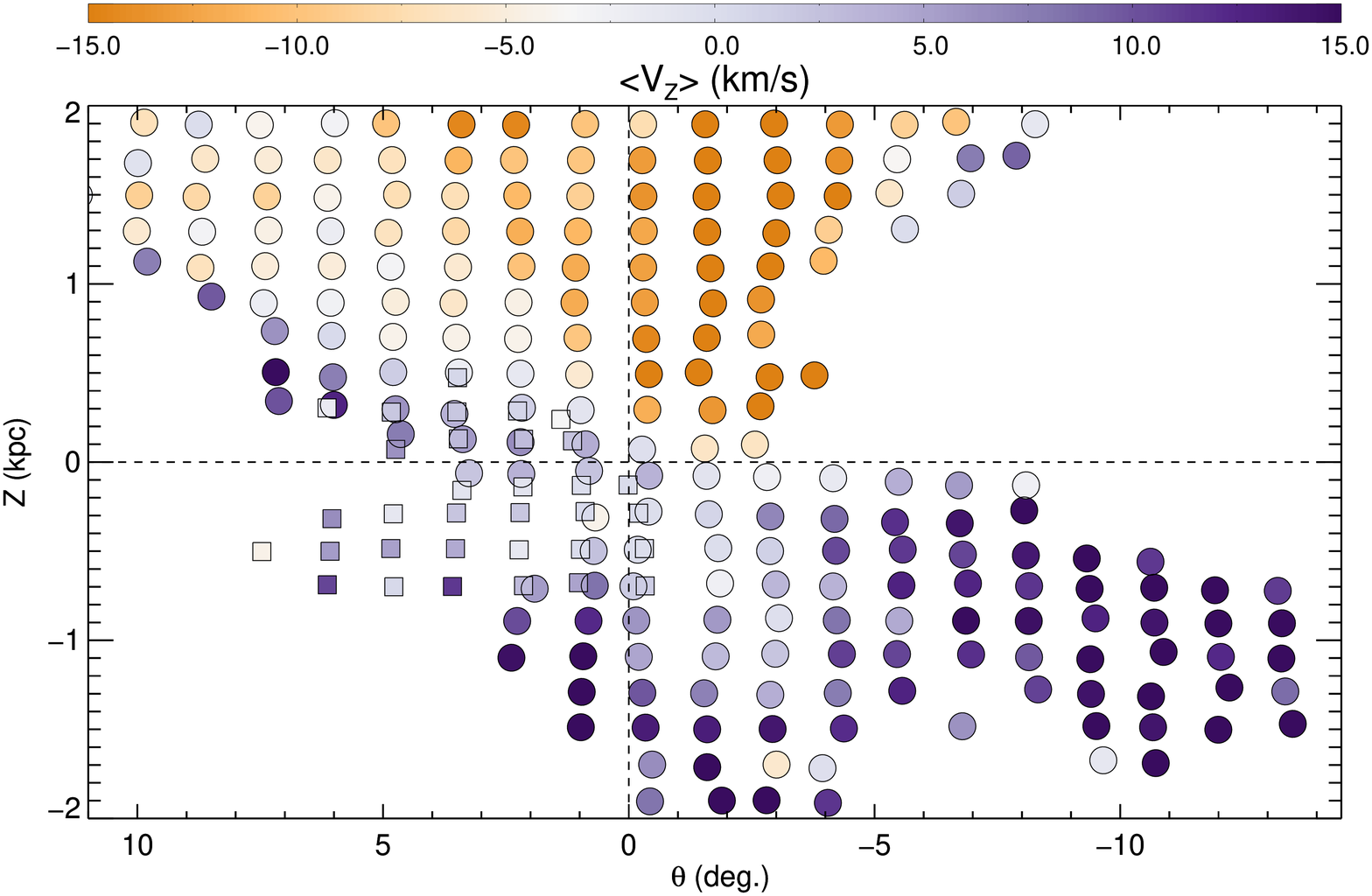}
 \includegraphics[width=0.49\textwidth, trim=0.5cm 0.0cm 0.5cm 0.5cm, clip]{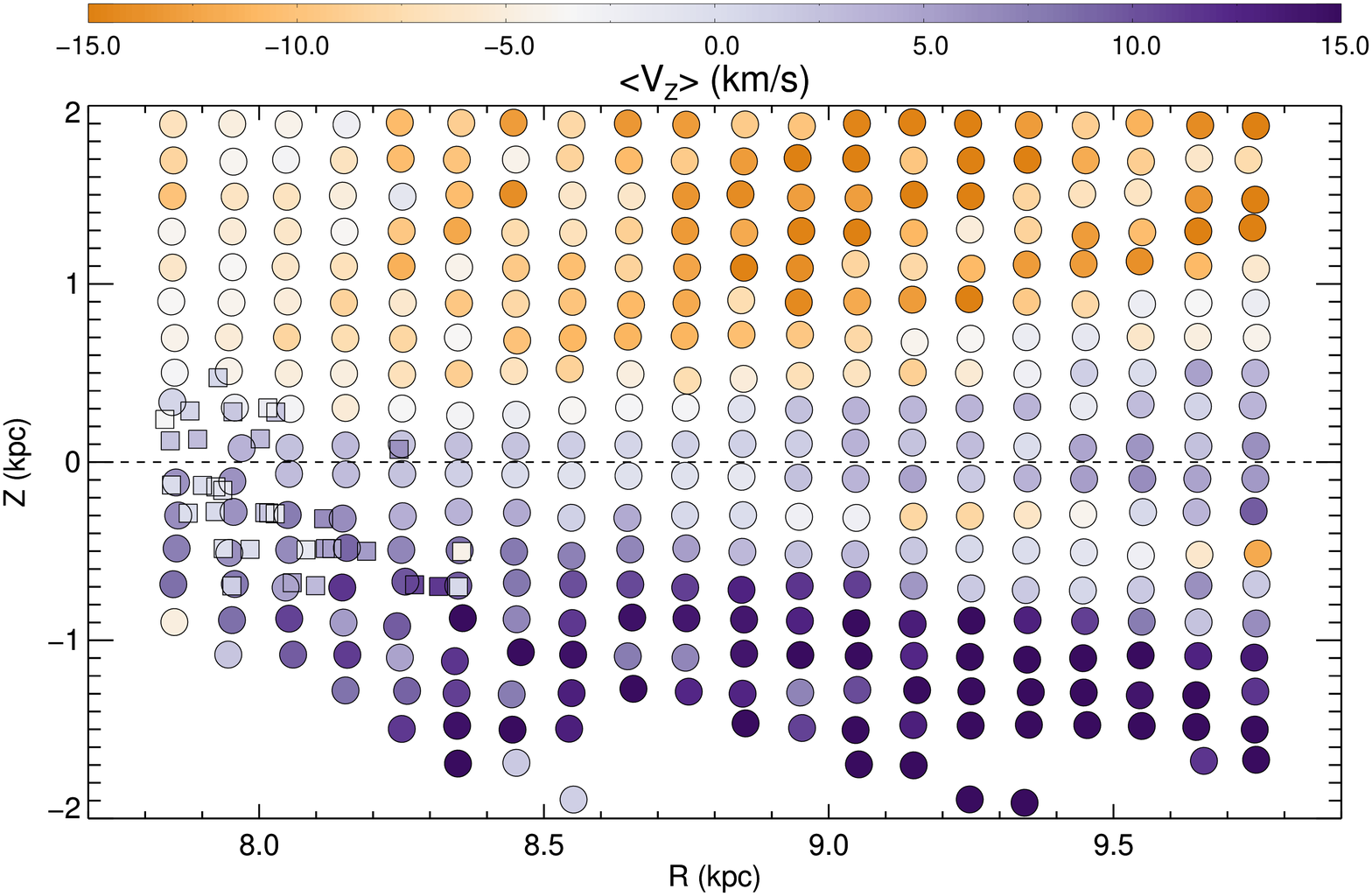}
 
 \caption{Radial ($V_R$, upper panels) and vertical ($V_Z$, lower panels) components of the Galactocentric velocities of stars between $7.8 < R_{\rm GC} < 9.8$~kpc as a function of positions in $Z$ and $\theta$ (left column) and $Z$, $R$ (right column). Circles denote LAMOST data, and small squares are derived from RAVE velocities. Each colored point represents the mean value of all stars within a bin 200~pc wide in $R$ and $Z$, and $1.3^\circ$ in $\theta$. All bins contain at least 50 stars, and some contain many thousands of stars. The dots are centered at the mean position of the stars within each subsample, and color encodes the mean $V_R$ or $V_Z$ according to the scale given by the color bar at the top. Apparent radial features in the right panels are artifacts consistent with $\sim20-30$\% errors in the distances.
 }
   \label{fig:vrvz_z_ang}
\end{center}
\end{figure}

We explore velocity structure within the cylindrical shell bounded by $7.8 < R_{\rm GC} < 9.8$ kpc and $-2.0 < Z < 2.0$~kpc as a function of position. Figure~\ref{fig:vrvz_z_ang} displays the mean $V_R$ (upper panels) and $V_Z$ (lower panels) velocities of stars binned spatially. 
The plots in the left column are centered on the Galactic anticenter ($Z = 0$~kpc, $\theta = 0^\circ
$), with the third Galactic quadrant to the left side, and the second quadrant to the right. Plots on the right display the velocity dependences with Galactocentric radius. We display the data as distinct dots so that it is obvious where 
LAMOST has observed, and which areas lack coverage.
Small filled squares in Figure~\ref{fig:vrvz_z_ang} represent the mean values for RAVE stars discussed in Section~\ref{sec:rave}. These data were 
selected with the same spatial criteria and error cuts as those from LAMOST. The RAVE data provide a 
small but important amount of additional coverage in $Z$ and $\theta$ that is not present in the LAMOST data.

If all of the relatively nearby stars in our sample were in disk-like orbits, one would not expect to see significant bulk motions in the radial or vertical directions. 
However, in the left column of Figure~\ref{fig:vrvz_z_ang}, the majority of the bins in both panels at $\theta<0^\circ$ show non-zero velocities. Furthermore, bins above the 
plane nearly all have large positive (outward) $\langle V_R \rangle$, while stars in the symmetric region below the Galactic plane have either very small 
or slightly negative ($\langle V_R \rangle < 0$~km~s$^{-1}$), inward-streaming motions. 
If the stars sampled disk-like rotational motion, we also should not see a change in $V_R$ across the Galactic anticenter. However, in this sample, 
the third Galactic quadrant ($\theta>0^\circ$) has velocities that are nearly zero in most bins, as expected. Then, near $\theta=0^\circ$, the velocities 
change to become significantly non-zero in the second Galactic quadrant.

The lower left panel of Figure~\ref{fig:vrvz_z_ang} shows that for the same stars at $\theta < 0^\circ$ that have opposite $\langle V_R \rangle$ velocities 
above and below the plane, the $\langle V_Z \rangle$ components also differ. Specifically, the stars above $Z\sim0.4$ kpc at $\theta < 0^\circ$ nearly all 
have downward ($\langle V_Z \rangle < 0$ km~s$^{-1}$) motions, while nearly all of the bins below the Galactic plane have $\langle V_Z \rangle > 
0$~km~s$^{-1}$.

The right column of Figure~\ref{fig:vrvz_z_ang} can be compared to contour plots of RAVE red clump giants seen in Figure~11 of \citet[note that these authors adopted $R_0 
= 8$~kpc rather than the value of 7.8~kpc we used]{wsb+13}. Indeed, for the region between $8 < R_{\rm GC} < 10$~kpc, the RAVE data exhibit similar 
asymmetries to those we see in LAMOST; namely, $\langle V_R \rangle > 0$~km~s$^{-1}$ above the midplane, and negative below the plane, while $
\langle V_Z \rangle$ shows stars moving downward in regions above the plane and upward for $Z < 0$~kpc. For regions outside the solar radius, the 
RAVE data of \citet{wsb+13} are nearly all at $0^\circ \lesssim \theta \lesssim 14.4^\circ$ and longitudes $l \gtrsim 225^\circ$. It is remarkable, then, that 
the same asymmetrical velocity features we see predominantly near the Galactic anticenter and in the second quadrant are also present in this RAVE 
sample.

\begin{figure}[!t]
\begin{center}
\includegraphics[width=0.65\textwidth, trim=0.0cm 0.0cm 0.0cm 0.0cm, clip]{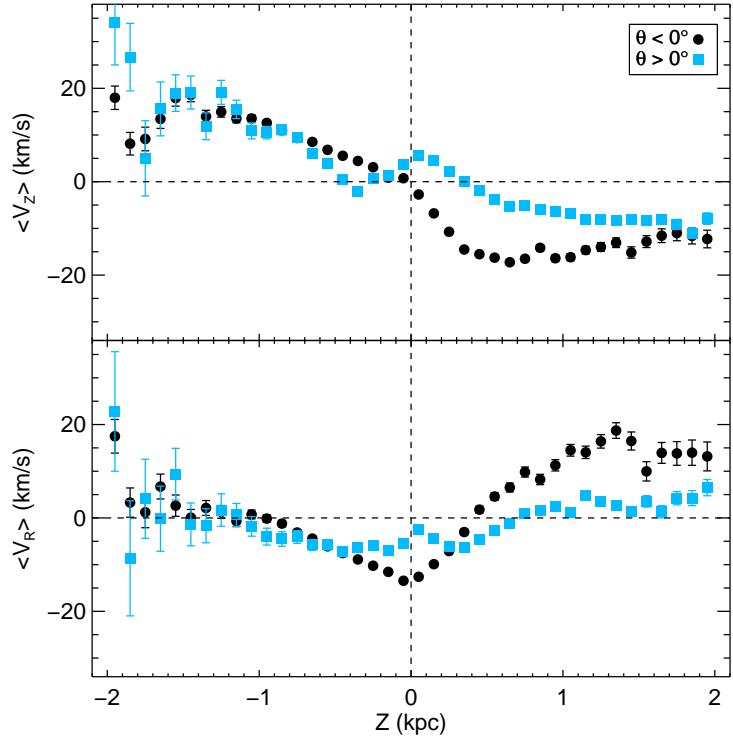}

\caption{Mean $V_Z$ and $V_R$ velocities of stars as a function of $Z$ in 0.1 kpc wide bins. Stars are separated into $\theta > 0^\circ$ (blue filled squares) and $\theta < 0^\circ$ (black filled circles) samples to examine differences between the second and third Galactic quadrants.  Error bars on each point represent the standard error on the mean; for points that average 
hundreds or thousands of stars, the error bars are smaller than the point size.}
\label{fig:vrvz_binned}
\end{center}
\end{figure}

To explore the azimuthal dependence (specifically, the difference between the second and third Galactic quadrants) of the vertical velocity asymmetries, 
we bin the data in a different way. Beginning with the entire data set of $\sim400,000$ stars, we separate them at the Galactic $X$-axis to create $\theta 
> 0^\circ$ and $\theta < 0^\circ$ samples. These subsets contain 212,264 and 178,632 stars, respectively. For each subset, we group 
the data into 0.1~kpc bins in $Z$ and calculate mean velocities $\langle V_R \rangle$ and $\langle V_Z \rangle$ for 
each bin. These are plotted in Figure~\ref{fig:vrvz_binned} as a function of $Z$. There are significant differences between the two samples, 
especially above the Galactic plane, but the overall trends are similar -- $\langle V_Z \rangle$ is positive below the plane and negative at $Z > 0$. In $
\langle V_R \rangle$, the $\theta > 0^\circ$ sample has smaller variations than the $\theta < 0^\circ$ subset, but shows a similar trend toward positive $
\langle V_R \rangle$ above the plane. There also appear to be small-scale wiggles (e.g., within $\pm0.5$~kpc of the plane in $\langle V_Z \rangle$) in 
the mean velocity that may be akin to the coherent wavelike motions seen by \citet{wgy+12} in SDSS data. 

We explored whether the unexpected motions in $\langle V_R \rangle$ and $\langle V_Z \rangle$ could be due to systematic errors in the kinematics and/or distances.
Measured line-of-sight velocities (which have at most only small systematic shifts; \citealt{lzz+12}) at the Galactic anticenter directly reflect the Galactocentric $V_R$ motions, but away from 
the anticenter the line-of-sight velocities give only a fraction of $V_R$. Because much of the 
velocity asymmetry we see is at $\left | Z \right | \gtrsim 1$~kpc, it may be sensitive to systematic shifts in the proper motions. 
Artificially shifting $\mu_{\alpha} \cos \delta$ primarily changes $V_R$ in trends parallel to the Galactic plane (i.e., the magnitude of the change induced by shifts in $\mu_{\alpha} \cos \delta$ is roughly constant at a given $\left | Z \right |$), 
while $\mu_{\delta}$ systematically shifts regions 
at $\left | Z \right | \gtrsim 0.5$~kpc in the same direction and by roughly the same amount above and below the plane. It is unlikely that systematic proper motion errors can account 
for the asymmetries that we see
both across the plane and across $\theta = 0^\circ$.
The $\langle V_Z \rangle$ component is primarily sensitive to the line-of-sight velocity above $\left | Z \right | \gtrsim 
1$~kpc. The vertical asymmetry in $\langle V_Z \rangle$ is thus robustly determined by spectroscopic measurements alone, though somewhat less secure at low latitudes, where the $Z$-component of Galactic velocities becomes sensitive to proper motion.
The effect of shifting the distances mostly shifts the mean velocities in each bin by roughly the same amount, leaving the differences that make up the asymmetry roughly the same.
Finally, by repeating the analysis on a subset of LAMOST spectra that have stellar parameters available, we verified that contamination by giant stars has little effect on our results.

\section{Conclusion}

We have shown an asymmetry in Galactocentric radial and vertical velocities across the midplane near the Galactic anticenter. F-type stars 
between $7.8 < R_{\rm GC} < 9.8$~kpc (i.e., within 2 kpc outside the Solar radius) and $\left | Z \right | < 2$~kpc are, on average, moving radially outward 
and downward toward the plane for positions above the midplane, and radially inward and upward toward the midplane for $Z < 0$. This persists over a 
large region in azimuth that is covered by LEGUE. The maximum velocity difference between regions above/below the 
plane is $\sim35$~km~s$^{-1}$ in $\langle V_Z \rangle$, and $\sim20$~km~s$^{-1}$ in $\langle V_R \rangle$. The asymmetry seems to weaken with 
azimuthal angle $\theta$ toward the third Galactic quadrant.

A comparison with results from RAVE \citep{wsb+13} shows that the same behavior is seen in the outer regions ($R_{\rm GC} \gtrsim 8.5$~kpc) 
probed by their data -- stars above the plane have net downward motions directed radially outward, while those below the plane are on average moving 
upward and in toward the Galactic center. At first glance, this would seem to be contradicted by the SDSS results of \citet{wgy+12}, in which the stars in 
their Figure~4 at $Z < 0$~kpc have $\langle V_Z \rangle < 0$~km~s$^{-1}$. These SDSS stars are located in the second Galactic quadrant (i.e., at $
\theta < 0^\circ$), where we find the strongest vertical asymmetry. Likewise, a similar negative $\langle V_Z \rangle$ is seen for stars below the plane 
from SDSS Stripe 82 (also mostly at $\theta < 0^\circ$) in the analysis by \citet[see their Figure~6]{swe12}, coupled with a positive $\langle V_R \rangle$ 
for these same stars. However, we argue that this is consistent with our findings. Both \citet{wgy+12} and \citet{swe12} restricted their 
stellar samples to $7 < R_{\rm GC} < 9$~kpc (with the Sun placed at $R_0 = 8$~kpc) and used subsets of SDSS stars that probe different regions of the disk than our data.
The RAVE data in Figure~11 of \citet{wsb+13} span $6.5 \lesssim R_{\rm GC} \lesssim 10.0$~kpc, and within this figure one can see that $\langle V_Z 
\rangle$ and $\langle V_R \rangle$ change significantly with position. Thus it is not surprising that the different (and smaller) volumes probed by the SDSS studies compared 
to our LAMOST sample should show different velocity trends than we find.

This velocity asymmetry is likely the result of a vertical disturbance of the disk by an 
external perturber \citep{wgy+12}, such as the Sagittarius dwarf galaxy (as suggested by \citealt{pbt+11, gmo+13}). The wavelike structure found in SDSS by \citet{gmo+12}, \citet{swe12}, and \citet
{wgy+12} is likely related to the substructure found in different volumes by \citet{wsb+13} in RAVE 
data, and by our study of LAMOST data at slightly larger radii. \citet{s12} showed that systematic errors in the line-of-sight velocities from SDSS could account for the north/south velocity asymmetry shown by \citet{wgy+12}. 
 However, the magnitude of the vertical velocity difference shown by LAMOST velocities in our study
suggests that such a systematic difference cannot be solely responsible for the observed substructure. Finally, we note that there is an overdensity seen in 
SDSS at $R_{\rm GC} \sim 9.5$~kpc, $Z \sim 0.6$~kpc, and $Y > 0$ by \citet[see their Figures 26 
and 27]{jib+08}, which may correspond to the regions in our study that show net outward and 
downward motions.

All of these pieces of evidence are merely suggestive; much larger data sets and more thorough 
analysis will be required to map the coherent velocity features in three dimensions. With more 
extensive mapping of kinematics in the extended solar neighborhood enabled by large 
spectroscopic surveys such as RAVE and LAMOST, and ultimately by the Gaia mission, further 
constraints can be placed on the merging and interaction history of the Milky Way.

\acknowledgements

We thank Brian Yanny for sharing the spectroscopic data from the \citet{wgy+12} study, and Mary 
Williams and Matthias Steinmetz for kindly sharing the RAVE data from \citet{wsb+13}. We are grateful to the anonymous referee for thoughtful and helpful comments. This work 
was supported by NSF grants AST 09-37523 and NSFC grant 10973015. Undergraduate research 
support was provided by the NASA/NY Space Grant fellowship and NSF grants AST 10-09670 and 
DMR 0850934. Guoshoujing Telescope (the Large Sky Area Multi-Object Fiber Spectroscopic 
Telescope, LAMOST) is a National Major Scientific Project built by the Chinese Academy of 
Sciences. Funding for the project has been provided by the National Development and Reform 
Commission. LAMOST is operated and managed by the National Astronomical Observatories, 
Chinese Academy of Sciences.

\bibliographystyle{apj}

\end{document}